\documentstyle[preprint,eqsecnum,aps]{revtex}

\def\half{{\textstyle{1\over2}}}

\def\pmb#1{\setbox0=\hbox{$#1$}%
\kern-.025em\copy0\kern-\wd0
\kern.05em\copy0\kern-\wd0
\kern-.025em\raise.0433em\box0}

\def\beq{\begin{equation}}
\def\eeq{\end{equation}}

\begin{document}

\def\footnoterule{\hrule width \hsize}
\def\footstrut{\baselineskip 16pt}

\skip\footins = 14pt 
\footskip     = 20pt 
\footnotesep  = 12pt 

\textwidth=6.5in
\hsize=6.5in
\oddsidemargin=0in
\evensidemargin=0in
\hoffset=0in

\textheight=9.5in
\vsize=9.5in
\topmargin=-.5in
\voffset=-.3in

\title{%
Two lectures on Two-Dimensional Gravity%
\footnotemark[1]}

\footnotetext[1]
{\baselineskip=16pt
This work is supported in part by funds provided by
the U.S.~Department of Energy (D.O.E.)
under contract \#DE-FC02-94ER40818. \hfil
MIT-CTP-2486 \hfil
November 1995\break}

\author{R. Jackiw}

\address{Center for Theoretical Physics,
Laboratory for Nuclear Science
and Department of Physics \\[-1ex]
Massachusetts Institute of Technology,
Cambridge, MA ~02139--4307}

\maketitle

\setcounter{page}{0}
\thispagestyle{empty}


\vspace*{\fill}

\begin{center}
In Memoriam: {\it Carlos Aragone}
\end{center}

\vspace*{\fill}

\begin{center}
LASSF II, Caracas, Venezuela, October 1995
\end{center}


\newpage

\widetext


When Carlos Aragone visited the physics community in Boston,
he became interested
in our on-going research
concerning lower-dimensional physics,
and he contributed
to our understanding
of gravitational models in (2+1) dimensions.
His death deprives us
of a valued research colleague;
Latin American compatriots will also
miss his organizational activities,
with which he endeavored
to keep physics
lively in Venezuela,
and more widely
in South America.

While Aragone was interested in and contributed to (2+1)-dimensional
gravity --- planar gravity --- I shall here speak on gravity in
(1+1)-dimensional space-time --- lineal gravity.  The purpose of
studying lower dimensional theories, and specifically lower dimensional
gravity, is to gain insight into difficult conceptional issues, which
are present and even more opaque in the physical (3+1)-dimensional
world.  Perhaps lessons learned in the lower-dimensional setting can be
used to explicate physical problems.  Moreover, if we are lucky, the
lower-dimensional theories can have a direct physical relevance to
modelling phenomena that is actually dynamically confined to the lower
dimensionality.  This is what happened with (2+1)-dimensional gravity:
gravitational physics in the presence of cosmic strings (infinitely
long, perpendicular to a plane) is adequately described planar gravity.
Indeed the recently discussed causality puzzles raised by ``Gott time
machines'' were resolved with the help of the lower-dimensional model
\cite{ref:1}.

In my first lecture I shall describe how gravity can be formulated as a
gauge theory, both classically and quantum mechanically.  In the second
lecture I shall discuss possible obstructions to quantizing
gravity-matter theory.

\section{Gauge Theories of Gravity}

A diffeomorphism-invariant gravity theory is obviously invariant against
transformations whose parameters are functions of space-time, just as in
a local gauge theory.  Consequently it has been long believed that gravity
theory can be formulated as a gauge theory, but in four dimensions it
has never been precisely clear how to do this.  On the other hand in
lower dimensions, it has been possible to give a clear and definite
prescription for a gauge theoretic formulation of the relevant gravity
theories.  Let me review the steps.

Step 1.   Formulate gravity, {\it not\/} in  terms of the metric
tensor $g_{\mu\nu}$, but rather in terms of Einstein-Cartan variables:
{\it Vielbeine\/} $e_\mu^a$ and spin-connections $\omega_\mu^{ab}$, which are
viewed    as independent     quantities --- the relation between them will
be given by an equation of motion.
[Notation:  Greek letters denote space-time components; Roman letters
denote components in a flat tangent space with metric
$\eta_{ab} = {\rm diag} (1, -1, \ldots)$;
$g_{\mu\nu} = e_\mu^a e_\nu^b \eta_{ab}$;
the anti-symmetric symbol $\epsilon^{ab\ldots}$ is normalized by
$\epsilon^{01\ldots} = 1$.]
In addition to $e_\mu^a$ and $\omega_\mu^{ab}$, it may be necessary to
use further variables, see below.   This formulation already gives
rise to a gauge theory of the local Lorentz group $G_L$, with
$\omega_\mu^{ab}$  being   the ``gauge potential,'' associated with the
Lorentz generator $J_{ab}, \omega_\mu \equiv \omega_\mu^{ab} J_{ab}$.
The {\it Vielbein\/} $e_\mu^a$ transforms contravariantly
under the Lorentz group --- it is {\it not\/} a potential.

Step 2.  Choose a gauge group $G$ that contains $G_L$, with generators
$Q_A$, which include the Lorentz generators $J_{ab}$, but also comprise
additional generators that are associated with variables other than
$\omega_\mu^{ab}$.  Specifically the {\it Vielbeine\/} are assocaited
with translations $P_a, ~ e_\mu = e_\mu^a P_a$, while other variables
(if they exist) are associated with other generators as needed.  In this
way one arrives at Lie-algebra valued potentials.
\beq
A_\mu = A_\mu^A Q_A = \omega_\mu + e_\mu + \ldots
\label{eq:1.1}
\eeq

Step 3.  Construct the gauge field curvature, in the usual way,
\beq
F_{\mu\nu} = \partial_\mu A_\nu - \partial_\nu A_\mu
+ [A_\mu, A_\nu]
\label{eq:1.2}
\eeq
where the commutator in (\ref{eq:1.2}) is evaluated from the Lie algebra
of the group.

Step 4.
It remains to construct a dynamical equation for these gauge fields,
such that the gauge field equation is recognized as the relevant
gravitational equation, when it is reexpressed in terms of the gravity
variables
$\omega_\mu^{ab}, e_\mu^a, \ldots$.
To this end,
we seek an action for the gauge variables,
which is gauge invariant,
and now we encounter
the first novelty of lower-dimensional gravity:
A gauge invariant action can indeed be given, but it is not of the
Yang-Mills paradigm $\langle \, F^{\mu\nu} , F_{\mu\nu} \, \rangle$,
rather use is made of novel structures available in
lower-dimensions.
[Here $\langle \, , \, \rangle$ denotes a symmetric, invariant bilinear
form on the Lie algebra; when the group is semi-simple this may be given
by the Killing-Cartan metric, for non semi-simple groups other invariants
are used, see below.]

(i) As is well known \cite{ref:2}, in (2+1) dimensions one can form the
Chern-Simons action.  When one uses the $ISO(2,1)$ Poincar\'e group one
arrives at a gauge theoretical formulation of Einstein
gravity.  Similarly using the $SO(3,1)$ or $SO(2,2)$, de Sitter or
anti-de Sitter groups leads to gravity with cosmological constant of one
or the other sign.  Finally conformal gravity arises from the $SO(3,2)$
group.  We do not here continue presenting the theory in this
dimensionality, but turn to the (1+1) dimensional case.

(ii)  When it comes to gravity in (1+1) dimensions, it is necessary to
{\it invent\/} a model, because Einstein's theory does not exist on a
line: the two-dimensional Einstein tensor
$R_{\mu\nu} - \half g_{\mu\nu} R$ vanishes identically;
the
Einstein-Hilbert action
$\int d^2 x \sqrt{-g} \, R$ is a surface term --- it is the topological Euler
invariant and does not lead to equations of motion.

A two-dimensional gravity model that has been widely studied by people
interested in string theory, conformal field theory and statistical
mechanics is Polyakov's induced gravity \cite{ref:3},
which is given by computing the
partition function for a massless field interacting with an external
gravitational field.  While this has engendered much research, nothing
particularly relevant to four-dimensional gravity has emerged ---
presumably because Polyakov's induced gravity action is non-local.

Even before Polyakov's proposal, Teitelboim and I
\cite{ref:4}
suggested that a way
to obtain non-trivial gravitational dynamics in (1+1) dimensions is to
introduce an additional gravitational variable, a world scalar Lagrange
multiplier field $\eta$, with which, together with the Riemann scalar,
one can construct non-trivial actions, {\it e.g.\/}
\beq
{\cal L}_1 = \sqrt{-g} \, \eta (R-\lambda)
\label{eq:1.3}
\eeq
where $\lambda$ is a cosmological constant.  Initially, only sporadic
research was performed on such models, but more recently it was realized
that similar
theories
arise in a limit of string theory.
Consequently many people, who
have become weary in the string program, have begun analyzing these
``string-inspired'' theories.  A particularly popular model is given by
the Callan-Giddings-Harvey-Strominger Lagrange density \cite{ref:5}
\beq
{\cal L}_2 = \sqrt{-g} \, (\eta R - \lambda)
\label{eq:1.4}
\eeq
which is like (\ref{eq:1.3}), except that the Lagrange
multiplier field $\eta$ does not multiply the cosmological constant.
[Actually CGHS present their model in terms of a ``dilaton'' field
$\phi$, related to $\eta$ by
$\eta = e^{-2\phi}$, and they use a rescaled metric tensor
$\bar{g}_{\mu\nu} = e^{2\phi} g_{\mu\nu} = g_{\mu\nu}/\eta$.
Thus in terms of the CGHS variables, their Lagrange density,
equivalent to (\ref{eq:1.4}),
reads
${\cal L}_{CGHS} = \sqrt{-\bar{g}} \, e^{-2\phi}
(\bar{R} + 4 \bar{g}^{\mu\nu} \partial_\mu \phi \partial_\nu \phi -
\lambda)$.]

A gauge theoretical formulation of these models uses the gauge curvature
tensor $F_{\mu\nu}^A$ for some appropriately chosen group, indexed by
$A$; evidently $F_{\mu\nu}^A$ transforms according to the adjoint
representation.  Next we take a multiplet of world scalar Lagrange
multipliers $\eta_A$, transforming according to the coadjoint
representation, with which an invariant Lagrange density can be
constructed, without use of a group metric.
\beq
{\cal L} = \half \eta_A \epsilon^{\mu\nu} F_{\mu\nu}^A
\label{eq:1.5}
\eeq
Gauge theories with a Lagrange density as in (\ref{eq:1.5}) are also
called B-F models, where ``B'' represents the Lagrange multiplier
multiplet.  One shows that with the
three-parameter
$SO(2,1)$ de Sitter or anti-de
Sitter groups, (\ref{eq:1.5}) produces the dynamics of ${\cal L}_1$
\cite{ref:6},
while use of the centrally extended Poincar\'e group $ISO(1,1)$, gives
rise to the dynamics of ${\cal L}_2$ \cite{ref:7}.  [The
(1+1)-dimensional, centrally extended Poincar\'e group possesses a
single Lorentz generator $J_{ab} = \epsilon_{ab} J$, two translation
generators $P_a,a=\left\{ 0,1 \right\}$, which unconventionally close on
a central element $I$: $[P_a, P_b] =
\epsilon_{ab} I$.  The spin connection $\omega^{ab} = \epsilon^{ab}
\omega_\mu$ is associated with $J$, the {\it Zweibein\/} $e_\mu^a$ with
$P_a$, and a further potential $a_\mu$ is needed for $I$.  Thus the
centrally extended Poincar\'e group contains four parameters; it is an
unconventional contraction of $SO(2,1)$.]

Step 5.  Finding classical solutions in the gauge theoretic formalism is
especially easy.   Varying $\eta_A$ in (\ref{eq:1.5}) gives
\beq
F_{\mu\nu} = 0
\label{eq:1.6}
\eeq
Consequently $A_\mu$ is a pure gauge,
\begin{mathletters}%
\label{eq:1.7}
\beq
A_\mu = U^{-1} \partial_\mu U
\label{eq:1.7a}
\eeq
where $U$ is an element of the group.  When the gauge $U=$ constant is
chosen, we get
\beq
A_\mu = 0
\label{eq:1.7b}
\eeq
\end{mathletters}%
Varying $A_\mu$ in (\ref{eq:1.5}) requires $\eta_A$ to be covariantly
constant.
\beq
D_\mu \eta = 0
\label{eq:1.8}
\eeq
In the gauge (\ref{eq:1.7b}) this means that $\eta$ is the constant
$\eta^0$, while more generally
\beq
\eta = U^{-1} \, \eta^0 \, U
\label{eq:1.9}
\eeq

The solution (\ref{eq:1.7b}) is especially provocative when it is
remembered that $A_\mu$ collects the gravitational variables
$\omega_\mu, ~ e_\mu^a, \ldots$
in terms of which the geometry of space-time is encoded.  Evidently when these
vanish, a space-time cannot be constructed.  But the problem is overcome
simply by choosing any gauge function $U$, such that (\ref{eq:1.7a}) is
non-singular and then the geometry can be reconstructed.  The {\it
invariant\/} characteristics of space-time are contained in the constant
past of $\eta_A$, as is seen from (\ref{eq:1.9}).  When a group metric
is available one may move indices and construct an invariant.
\beq
C=<\eta, \eta> = \eta_A \eta^A
\label{eq:1.10}
\eeq
[For $SO(2,1)$ the metric is the Cartan-Killing metric; for centrally extended
$ISO(1,1)$, which is not semi-simple, the Cartan-Killing bilinear form
is singular and cannot serve as a metric, but an alternate metric can be
constructed, so that $\eta_A \eta^A = \eta_a \eta^a - 2 \eta_2 \eta_3$.
Because this group is solvable, there also exists an invariant vector
$V^A$, so that $V^A \eta_A = \eta_3$ is also an invariant.]
Note that the extended $ISO(1,1)$ model $\approx$ CGHS theory does {\it
not\/} possess a cosmological constant in its gauge theoretical
formulation:
$\lambda$ emerges as a solution to the equations motion, which require,
{\it inter alia\/}, that
$\eta'_3= 0$, $\eta_3=\lambda$. (The dash signifies differentiation with
respect to the spatial variable $\sigma$.)

Step 6:
Quantization of pure gravity is effected by taking the gravitational
Lagrange density ${\cal L}_g$ to be proportional to (\ref{eq:1.5}).
\beq
(4\pi G) {\cal L}_g =
\eta_A F_{01}^A
= \eta_A \dot{A}_1^A
+ (4\pi G) A_0^A  G_A
\label{eq:1.11}
\eeq
\beq
G_A = {1\over (4\pi G)} (D_1 \eta)_A
\label{eq:1.12}
\eeq
[$G$ is ``Newton's'' gravitational constant,
the over-dot signifies differentiation with respect to time $t$.]
The last equality in (\ref{eq:1.11}) follows from the previous after a
spatial integration by parts.  The phase-space form of (\ref{eq:1.11})
shows that $\eta_A$ and $A_1^A$ are a canonical pair: momentum and
coordinate respectively, satisfying the quantum commutator
\beq
{}[A_1^A(\sigma), \eta_B (\tilde{\sigma})] = 4\pi G \, \delta_B^A \, i
\delta(\sigma-\tilde{\sigma})
\label{eq:1.13}
\eeq

The quantities $G_A$ in (\ref{eq:1.12}) are the ``Gauss Law''
constraints, whose algebra can be computed with the help of
(\ref{eq:1.13}) and is found to follow the Lie algebra of the relevant
group, with structure constants ${f_{AB}}^C$.
\begin{eqnarray}
{}[Q_A, Q_B] &=& {f_{AB}}^C Q_C
\label{eq:1.14} \\
{}[G_A(\sigma), G_B(\tilde{\sigma})] &=& i \, {f_{AB}}^C G_C (\sigma)
\delta(\sigma-\tilde{\sigma})
\label{eq:1.15}
\end{eqnarray}
These constraints are first-class, they can be imposed on states
\beq
G_A(\sigma) | \, \Psi \, \rangle = 0 \Rightarrow
(\eta'_A + {f_{AB}}^C A_1^B \eta_C ) | \, \Psi \, \rangle = 0
\label{eq:1.16}
\eeq
Although no group metric is needed to present (\ref{eq:1.16}), when a
group metric does exist (as in our two examples)
$f_{ABC}$ is totally anti-symmetric and contracting (\ref{eq:1.16}) with
$\eta^A$ leaves
\beq
\eta^A \eta'_A | \, \Psi \, \rangle =
{\half} (\eta^A \eta_A)' | \, \Psi \, \rangle = 0
\label{eq:1.17}
\eeq
This means that physical states satisfying (\ref{eq:1.17})
possess support only for the constant portion of the invariant
$\eta^A \eta_A$.
Moreover, when there exists an invariant vector $V^A$,
{\it e.g.\/} when the group is the centrally extended $ISO(1,1)$, and
$V^A {f_{AB}}^C = 0$,
it follows from (\ref{eq:1.16}) that
\beq
V^A \eta'_A | \, \Psi \, \rangle = 0
\label{eq:1.18}
\eeq
so that $| \Psi \rangle$ has support only for the constant part of the
invariant $V^A \eta_A$.

The abstract states $| \Psi \rangle$ may be
explicitly presented in the Schr\"odinger representation with momentum
polarization, such that states are wave functionals of $\eta_A$ and
$A_1^A$ is realized as $4\pi G \, i {\delta \over \delta \eta_A}$.
Thus Eq.~(\ref{eq:1.16}) becomes
\beq
(\eta'_A + 4 \pi G \, i {f_{AB}}^C \, \eta_C \, {\delta \over \delta
\eta_B})
\, \Psi(\eta) = 0
\label{eq:1.19}
\eeq

Before discussing solutions to this equation, let me record a peculiar
and not-very-well-known fact about gauge theories in the Schr\"odinger
representation.  Let us recall that on the canonical variables,
coordinate ${A}$ and momentum $\Pi$, gauge transformations act as
\begin{mathletters}%
\begin{eqnarray}
A^U &=& U^{-1} A U + U^{-1} \, dU
\label{eq:1.20a} \\
\Pi^U &=& U^{-1} \Pi U
\label{eq:1.20b}
\end{eqnarray}
\end{mathletters}%
The above holds for Yang-Mills theories (where $\Pi$ is the electric
field)
and B-F theories where $\Pi$ is $\eta$ --- both are gauge covariant.
(It does {\it not\/} hold in theories with a Chern-Simons term, because
then $\Pi$ is not gauge covariant.)   In the coordinate representation,
where wave functionals $\Phi$ depend on $A$, Gauss' law ensures that
they are gauge invariant (we ignore the topological vacuum angle).
\beq
\Phi(A^U) = \Phi(A)
\label{eq:1.21}
\eeq
%
%
The question then arises, how do the momentum space wave functionals
$\Psi(\Pi)$ respond to gauge transformations?  The answer is gotten by
considering a functional Fourier transform representation for
$\Psi(\Pi)$.
\begin{mathletters}%
\beq
\Psi(\Pi) = \int {\cal D} A e^{-i \int \langle \Pi,A \rangle} \Phi(A)
\label{eq:1.22a}
\eeq
Using (\ref{eq:1.21})
and performing various changes of integration variabels, with unit
Jacobian, leaves
\begin{eqnarray}
\Psi(\Pi)
&=&
\int {\cal D} A e^{-i \int \langle \Pi,A \rangle}
\Phi(U^{-1} A U + U^{-1} dU )
\nonumber\\
&=&
\int {\cal D} A e^{-i \int \langle \Pi,U \! A U^{-1} \rangle}
\Phi(A + U^{-1} dU )
\nonumber\\
&=&
\int {\cal D} A e^{-i \int \langle U \Pi U^{-1}, A - U^{-1} dU \rangle}
\Phi(A)
\nonumber\\
&=&
e^{i \int \langle \Pi dU \, U^{-1} \rangle}
\Psi(\Pi^U)
\label{eq:1.22b}
\end{eqnarray}
\end{mathletters}%
Thus, unlike in the coordinate polarization, with the momentum
polarization wave functionals are {\it not\/} gauge invariant, rather
they satisfy the 1-cocycle transformation law \cite{ref:8}.
\beq
\Psi(\Pi^U) = e^{-i \int \langle \Pi, dU \, U^{-1} \rangle}
\, \Psi(\Pi)
\label{eq:1.23}
\eeq
One may extract a phase from $\Psi$ and work with a gauge invariant
functional $\psi$, by defining
\beq
\Psi(\Pi) = e^{-i \Omega(\Pi)} \, \psi(\Pi)
\label{eq:1.24}
\eeq
$\psi(\Pi)$ is gauge invariant, consistent with (\ref{eq:1.23}),
provided $\Omega$ satisfies
\beq
\Omega(U^{-1} \Pi U) - \Omega(\Pi) = \int \langle \Pi, dU \, U^{-1} \rangle
\label{eq:1.25}
\eeq
In particular, for the B-F theories that we have been considering in one
spatial dimension, where $\Pi$ is ${1 \over 4 \pi G} \eta$, we seek a
functional $\Omega(\eta)$, such that
\begin{mathletters}%
\beq
\Omega(U^{-1} \eta U ) - \Omega (\eta) =
{1 \over  4 \pi G} \int \langle \eta, dU \, U^{-1} \rangle
\label{eq:1.26a}
\eeq
The solution to this is
\beq
\Omega(\eta) = {1 \over 4 \pi G} \int K(\eta) \, d \sigma
\label{eq:1.26b}
\eeq
\end{mathletters}%
where $K(\eta) d \sigma$ is the Kirillov-Kostant 1-form, evaluated on the
co-adjoint orbit of the relevant group.

Returning now to our specific gravity models, we conclude by recording
the solution to the constraints for the CGHS model, based on centrally
extended $ISO(1,1)$.
With the explicit structure constants,
the four equations of (\ref{eq:1.19}) are solved by
\beq
\Psi(\eta) = e^{-i\Omega}
\delta(\eta'_3)
\delta \left(
(\eta_a \eta^a - 2 \eta_2 \eta_3)' \right) \psi
\nonumber
\eeq
where $\Omega = {1\over 8\pi G \lambda} \int \epsilon^{ab} d\eta_a \eta_b$
is the relevant Kirillov-Kostant form.
The functional $\delta$ functions ensure that $\psi$ depends only on the
constant parts of the two invariants $V^A\eta_A = \eta_3$ and $\eta_A \eta^A =
\eta_a
\eta^a - 2 \eta_2 \eta_3$; $\lambda$ is (the constant part of) $\eta_3$
--- it is the cosmological constant \cite{ref:9}.


\newpage

\section{Obstructions to Quantizing Gravity Theory
and Quantal Modification of Wheeler De Witt Equation}


In a canonical, Hamiltonian approach to quantizing a theory with local
symmetry --- a theory that is invariant against transformations whose
parameters are arbitrary functions on space-time ---  there occur
constraints, which are imposed on physical states.   Typically these
constraints correspond to time components of the Euler-Lagrange
equations, and familiar examples arise in gauge theories.   The time
component of the gauge field equation is the Gauss law.
\beq
G_A \equiv {\bf D} \cdot {\bf E}^A - \rho^A = 0
\label{eq:2.1}
\eeq
Here ${\bf E}^A$ is the (non-Abelian) electric field, $\rho^A$ the
matter charge density, and ${\bf D}$ denotes the gauge-covariant
derivative.  When expressed in terms of canonical variables, $G_A$ does
{\it not\/} involve time-derivatives --- it depends on canonical
coordinates and momenta, which we denote collectively by the symbols $X$
and $P$ respectively  ($X$ and $P$ are fields defined at fixed time)
$: ~ G_A = G_A (X,P)$.   Thus in a Schr\"odinger representation for the
theory, the Gauss law condition on physical states
\beq
G_A (X,P) | \, \psi \, \rangle = 0
\label{eq:2.2}
\eeq
corresponds to a (functional)  differential equation that the state
functional $\Psi(X)$ must satisfy.
\beq
G_A \left( X, {1\over i} {\delta \over \delta X} \right)
\, \Psi(X) = 0
\label{eq:2.3}
\eeq
In fact, Eq.~(\ref{eq:2.3}) represents an infinite number of equations,
one for each spatial point ${\bf r}$, since $G_A$ is also the generator
of the local symmetry:
$G_A = G_A({\bf r})$.
Consequently, questions of consistency (integrability) arise, and these
may be examined by considering the commutator of two constraints.
Precisely because the $G_A$ generate the symmetry transformation,
one expects their commutator to follow the Lie algebra with structure
constants ${f_{AB}}^C$.
\beq
\left[ G_A ({\bf r}) , G_B (\tilde{\bf r}) \right]
= i \, {f_{AB}}^C \, G_C ({\bf r}) \, \delta({\bf r} - \tilde{\bf r})
\label{eq:2.4}
\eeq
If (\ref{eq:2.4}) holds, the constraints are consistent --- they are first
class --- and the constraint equations are integrable, at least locally.

However, it is by now well-known that Eq.~(\ref{eq:2.4}), which {\it
does\/} hold classically with Poisson bracketing, may acquire a quantal
anomaly.  Indeed when the matter charge density is constructed from
fermions of a definite chirality, the Gauss law algebra is modified by
an extension --- a Schwinger term --- the constraint equations become
second-class and Eq.~(\ref{eq:2.3}) is inconsistent and cannot be solved.
We call such gauge theories ``anomalous.''

This does not mean that a quantum theory cannot be constructed from an
anomalous gauge theory.  One can adopt various strategies for overcoming
the obstruction, but these represent modifications of the original
model.  Moreover, the resulting quantum theory possesses physical
content that is very far removed from what one might infer by studying
the classical model.  All this is explicitly illustrated by the
anomalous chiral Schwinger model, whose Gauss law is obstructed, while a
successful construction of the quantum theory leads to massive
excitations, which cannot be anticipated from the un-quantized equations
\cite{ref:10}.

With these facts in mind, we turn now to gravity theory, which obviously
is invariant against local transformations that redefine coordinates of
space-time.

Indeed over the years there have been many attempts to describe gravity
in terms of a gauge theory.  That program is entirely successful in
three- and two-dimensional space-time, where gravitational models are
formulated in terms of Einstein--Cartan variables (spin-connection, {\it
Vielbein}) as non-Abelian gauge theories, based not on the Yang-Mills
paradigm, but rather on topological Chern-Simons and B-F structures.

But even remaining with the conventional metric-based formulation, it
is recognized that the time components of Einstein's equation comprise
the constraints.
\beq
{1\over   8\pi G}  \left( R_\nu^{\,0} -
{\textstyle{1\over2}} \delta_\nu^{\,0} R \right)
- T_\nu^{\,0} = 0
\label{eq:2.5}
\eeq
The gravitational  part is the time component of the Einstein tensor
$R^\mu_{\nu} - {1\over2} \delta^\mu_{\nu} R$;
weighted by Newton's constant $G$, this equals the time component of the
matter energy-momentum tensor, $T^{\mu}_{\nu}$.   In the quantized
theory, the collection of canonical operators on the left side in
(\ref{eq:2.5}) annihilates physical states.  The resulting equations may be
presented as
\begin{eqnarray}
{\cal E} \, | \, \psi \, \rangle &=& 0 ~~,
\label{eq:2.6} \\
{\cal P}_i \, | \, \psi \, \rangle &=& 0 ~~,
\label{eq:2.7}
\end{eqnarray}
where ${\cal E}$ is the energy constraint
\beq
{\cal E} = {\cal E}^{\rm\,gravity} + {\cal E}^{\rm\,matter} ~~,
\label{eq:2.8}
\eeq
and ${\cal P}_i$ is the momentum constraint.
\beq
{\cal P}_i = {\cal P}_i^{\rm\,gravity} + {\cal P}_i^{\rm\,matter}
\label{eq:2.9}
\eeq
Taking for definiteness matter to be described by a massless, spinless
field $\varphi$, with canonical momentum $\Pi$, we have
\begin{eqnarray}
{\cal E}^{\rm\,matter} &=&
{\textstyle{1\over2}} \left( \Pi^2 + \gamma \,
\gamma^{ij} \, \partial_i \, \varphi \, \partial_j \, \varphi \right)
\label{eq:2.10} \\
{\cal P}_i^{\rm\,matter} &=& \partial_i \, \varphi \, \Pi
\label{eq:2.11}
\end{eqnarray}
Here $\gamma_{ij}$ is the spatial metric tensor;
$\gamma$, its determinant;
$\gamma^{ij}$, its inverse.

The momentum constraint in Eq.~(\ref{eq:2.7}) is easy to unravel.
In a Schr\"odinger representation, it requires that
$\Psi(\gamma_{ij}, \varphi)$
be a functional of the canonical field variables
$\gamma_{ij},\varphi$
that is invariant against reparameterization of the spatial
coordinates and such functionals are easy to construct.

Of course it is (\ref{eq:2.6}), the Wheeler-DeWitt equation, that is
highly non-trivial and once again one asks about its consistency.  If
the commutators of ${\cal E}$ with ${\cal P}$ follow their Poisson
brackets one would expect
that the following algebra holds.
\begin{mathletters}%
\label{eq:2.12all}
\begin{eqnarray}%
\left[ {\cal P}_i ({\bf r}), {\cal P}_j (\tilde{\bf r}) \right]
&=& i {\cal P}_j ({\bf r}) \, \partial_i \, \delta({\bf r} - \tilde{\bf r})
+ i {\cal P}_i (\tilde{\bf r}) \, \partial_j \, \delta({\bf r} - \tilde{\bf r})
\label{eq:2.12a} \\
\left[ {\cal E} ({\bf r}), {\cal E} (\tilde{\bf r}) \right]
&=& i \left( {\cal P}^i ({\bf r}) + {\cal P}^i (\tilde{\bf r}) \right)
\, \partial_i \, \delta({\bf r} - \tilde{\bf r})
\label{eq:2.12b} \\
\left[ {\cal E} ({\bf r}), {\cal P}_i (\tilde{\bf r}) \right]
&=& i \left( {\cal E} ({\bf r}) + {\cal E} (\tilde{\bf r}) \right)
\, \partial_i \, \delta({\bf r} - \tilde{\bf r})
\label{eq:2.12c}
\end{eqnarray}%
\end{mathletters}%
Here ${\cal P}^i \equiv \gamma \, \gamma^{ij} \, {\cal P}_j$.
If true,
Eqs.~(\ref{eq:2.12all})
would demonstrate the consistency of the
constraints, since they appear first-class.
Unfortunately, establishing
(\ref{eq:2.12all})
in the quantized theory is highly problematical.
First of all there is the issue of operator ordering in the gravitational
portion of ${\cal E}$ and ${\cal P}$.  Much has been said about this,
and I shall not address that difficulty here.

The problem that I want to call attention to is the very likely
occurrence of an extension in the $[{\cal E}, {\cal P}_i]$ commutator
(\ref{eq:2.12c}).  We know that in flat space, the commutator between the
matter energy and momentum densities possesses a triple derivative
Schwinger term \cite{ref:11}.
There does not appear any known mechanism
arising from the gravity variables
that would effect a cancelation of this obstruction.

A definite resolution of this question in the full quantum theory is out
of reach at the present time.  Non-canonical Schwinger terms can be
determined only after a clear understanding of the singularities in the
quantum field theory and the nature of its Hilbert space are in hand,
and this is obviously lacking for four-dimensional quantum gravity.

Faced with the impasse, we turn to a gravitational model in
two-dimensional space-time
--- a {\it lineal\/} gravity theory ---
where the calculation can be carried to a
definite conclusion: an obstruction does exist and the model is
anomalous.  Various mechanisms are available to overcome the anomaly,
but the resulting various quantum theories
are inequivalent and bear little resemblance to the classical model.

In two dimensions, Einstein's equation is vacuous because
$R_\nu^\mu = {1\over2} \delta^\mu_\nu R$;
therefore gravitational dynamics has to be invented afresh.   The models
that have been studied recently posit local dynamics for the ``gravity''
sector, which involves as variables the metric tensor and an additional
world scalar (``dilaton'' or Lagrange multiplier)  field.  Such
``scalar-tensor'' theories, introduced a decade ago
\cite{ref:4},
are obtained by
dimensional reduction from higher-dimensional Einstein theory
\cite{ref:4,ref:12}.
They should be contrasted with models where quantum fluctuations of
matter variables induce gravitational dynamics
\cite{ref:3},
which therefore are
non-local and do not appear to offer any insight into the questions
posed by the physical, four-dimensional theory.

The model we study is the so-called ``string-inspired dilaton gravity''
-- CGHS theory \cite{ref:5}.
The gravitational action involves the metric
tensor $g_{\mu\nu}$, the dilaton field
$\phi, ~\eta \equiv e^{-2\phi}$,
and a cosmological
constant $\lambda$.  The matter action describes the coupling of a
massless, spinless field $\varphi$.
\begin{eqnarray}
I_{\rm\,gravity} &=& \int d^2x \, \sqrt{-g} \, \eta
\left( R - \lambda \right)
\label{eq:2.13} \\
I_{\rm\,matter} &=& {\textstyle{1\over2}} \int d^2 x \,
\sqrt{-g} \, g^{\mu\nu} \, \partial_\mu \varphi \partial_\nu \varphi
\label{eq:2.14}
\end{eqnarray}
The total action is the sum of (\ref{eq:2.13}) and (\ref{eq:2.14}),
weighted by ``Newton's'' constant $G$:
\beq
I = {1\over 4\pi G} ~ I_{\rm\,gravity} + I_{\rm\,matter}
\label{eq:2.15}
\eeq

Before embarking on the quantal analysis let us remark that classically
the theory can be solved completely.  Indeed, varying $\eta$ shows that
$R$ vanishes, space-time is flat, and $I_{\rm\,matter}$ describes the
flat-space motion of free massless scalar field, to be sure in a diffeomorphism
invariant fashion.  One would hope to regain this
simple dynamics after quantization.  For point particles, where
$I_{\rm\,matter} = -m
{\displaystyle\int}
\sqrt{g_{\mu\nu} d x^\mu d x^\nu}$,
the quantization is successful: one finds a diffeomorphism invariant,
quantum description of free particles \cite{ref:13}.  With matter fields,
the quantum development encounters difficulties.

In fact this theory can be given a gauge-theoretical ``B-F''
description
based on the
centrally extended
Poincar\'e group
in (1+1) dimensions \cite{ref:14}.
This formulation aided us immeasurably    in the subsequent
analysis/transformations.
However, I shall not discuss this here,
because in retrospect it proved possible
to carry the analysis forward within the
metric formulation (\ref{eq:2.13}--\ref{eq:2.15}).

After a remarkable sequence of redefinitions and canonical
transformations on the dynamical variables in
(\ref{eq:2.13}--\ref{eq:2.15}),
one can present $I$ in terms
of a first-order Lagrange density ${\cal L}$ that is a sum of
quadratic terms \cite{ref:14}.
\begin{eqnarray}
{\cal L} &=& \pi_a \dot{r}^a + \Pi \dot{\varphi}
- \alpha {\cal E} - \beta {\cal P}
\label{eq:2.16} \\
{\cal E} &=& - {\textstyle{1\over2}}
\left(
{\textstyle{1\over \Lambda}} \pi^a \pi_a + \Lambda {r^a}' {r_a}' \right)
+ {\textstyle{1\over2}} \left( \Pi^2 + {\varphi'}^2 \right)
\label{eq:2.17} \\
{\cal P} &=& - {r^a}' \pi_a  - \varphi' \Pi
\label{eq:2.18}
\end{eqnarray}
I shall not derive this, but merely explain it.
The index ``a'' runs over flat 2-dimensional $(t,\sigma)$ space, with
signature $(1,-1)$.  Dot (dash) signify differentiation with respect to
time $t$ (space $\sigma$).  The four variables
$\left\{ r^a, \alpha, \beta \right\}$ correspond to the four
gravitational variables $(g_{\mu\nu}, \eta)$, where only $r^a$ is
dynamical with canonically  conjugate momentum $\pi_a$, while $\alpha$
and $\beta$ act as Lagrange multipliers.  Notice that regardless of the
sign $\Lambda \equiv \lambda / 8 \pi G$, the gravitational
contribution to ${\cal E}$, is quadratic with indefinite sign.
\begin{mathletters}%
\label{eq:2.19}
\begin{eqnarray}
{\cal E}^{\rm\,gravity} &=&
- {\textstyle{1\over2}} \left(
{\textstyle{1\over\Lambda}} \pi^a \pi_a +
\Lambda {r^a}' {r_a}' \right) \nonumber \\
&=& - {\textstyle{1\over2}} \left(
{\textstyle{1\over\Lambda}} (\pi_0)^2
- {\textstyle{1\over\Lambda}} (\pi_1)^2
+ \Lambda ({r^0}')^2
- \Lambda ({r^1}')^2 \right) \nonumber\\
&=& - {\cal E}_0 + {\cal E}_1
\label{eq:2.19a} \\
{\cal E}_0 &=& {\textstyle{1\over2}} \left(
{\textstyle{1\over\Lambda}}
(\pi_0)^{2} + \Lambda ({r^0}')^2 \right)
\label{eq:2.19b} \\
{\cal E}_1 &=& {\textstyle{1\over 2}}
\left( {\textstyle{1\over\Lambda}}
(\pi_1)^{2} + \Lambda ({r^1}')^2 \right)
\label{eq:2.19c}
\end{eqnarray}
\end{mathletters}%
On the other hand, the gravitational contribution to the momentum does
not show alteration of sign.
\begin{mathletters}%
\label{eq:2.20}
\begin{eqnarray}
{\cal P}^{\rm\,gravity} &=& - {r^a}' \pi_a \nonumber\\
                        &=& - {r^0}' \pi_0  - {r^1}' \pi_1 \nonumber\\
                        &=& {\cal P}_0 + {\cal P}_1
\label{eq:2.20a} \\
{\cal P}_0 &=& - {r^0}' \pi_0
\label{eq:2.20b} \\
{\cal P}_1 &=& - {r^1}' \pi_1
\label{eq:2.20c}
\end{eqnarray}
\end{mathletters}%
One may understand the relative negative sign        between the two
gravitational    contributors $(a=0,1)$ as follows.   Pure   metric
gravity   in two space-time dimensions is described by three functions
collected   in $g_{\mu\nu}$.  Diffeomorphism invariance involves 2
functions, which reduce the number of variables by $2\times2$,
{\it i.e.\/} pure gravity has $3-4=-1$ degrees of freedom.  Adding the
dilaton $\phi$ gives a net number of $-1+1=0$, as in our final
gravitational Lagrangian.

The matter contribution    is the conventional expression for massless
and spinless fields:
\begin{eqnarray}
{\cal E}^{\rm\,matter} &=& {\textstyle{1\over2}} (\Pi^2 + {\varphi'}^2)
\label{eq:2.21} \\
{\cal P}^{\rm\,matter} &=& - \varphi' \, \Pi
\label{eq:2.22}
\end{eqnarray}
It is with the formulation in
Eqs.~(\ref{eq:2.16})--(\ref{eq:2.22})
of the theory (\ref{eq:2.13})--(\ref{eq:2.15})
that we embark upon the various quantization procedures.

The transformed theory appears very simple:  there are three independent
dynamical fields
$\left\{ r^a, \varphi \right\}$
and together with the canonical momenta
$\left\{ \pi_a, \Pi \right\}$
they lead to a quadratic Hamiltonian, which has
no interaction terms among the three.  Similarly, the momentum comprises
non-interacting terms.  However, there remains a subtle ``correlation
interaction'' as a consequence of the constraint that ${\cal E}$ and
${\cal P}$ annihilate physical states, as follows from varying the
Lagrange multipliers $\alpha$ and $\beta$ in (\ref{eq:2.16}).
\begin{eqnarray}
{\cal E} \, | \, \psi \, \rangle  &=& 0
\label{eq:2.23} \\
{\cal P} \, | \, \psi \, \rangle  &=& 0
\label{eq:2.24}
\end{eqnarray}
Thus, even though ${\cal E}$ and ${\cal P}$ each are sums of
non-interacting variables, the physical states
$| \, \psi \, \rangle$ are not      direct products of states for the
separate degrees of freedom.  Note that Eqs.~(\ref{eq:2.23}), (\ref{eq:2.24})
comprise the entire physical content of the theory.   There is no need
for any further ``gauge fixing'' or ``ghost'' variables --- this is the
advantage of the Hamiltonian formalism.

As in four dimensions, the momentum constraint
(\ref{eq:2.24}) enforces invariance
of the state functional $\Psi(r^a, \varphi)$ against spatial coordinate
transformations, while the energy constraint
(\ref{eq:2.23})
--- the Wheeler-DeWitt
equation in the present lineal gravity context --- is highly non-trivial.

Once again one looks to the algebra of the constraints
to check consistency.
The reduction of
(\ref{eq:2.12all})
to one spatial dimension leaves (after the identification
${\cal P}_i \to - {\cal P}, \gamma \gamma^{ij} \to 1$),
\begin{mathletters}%
\label{eq:2.25}
\begin{eqnarray}
i [ {\cal P}(\sigma), {\cal P}(\tilde{\sigma}) ] &=&
\left( {\cal P}(\sigma) + {\cal P}(\tilde{\sigma}) \right)
\, \delta'(\sigma - \tilde{\sigma})
\label{eq:2.25a} \\
i [ {\cal E}(\sigma), {\cal E}(\tilde{\sigma}) ] &=&
\left( {\cal P}(\sigma) + {\cal P}(\tilde{\sigma}) \right)
\, \delta'(\sigma - \tilde{\sigma})
\label{eq:2.25b} \\
i [ {\cal E}(\sigma), {\cal P}(\tilde{\sigma}) ] &=&
\left( {\cal E}(\sigma) + {\cal E}(\tilde{\sigma}) \right)
\, \delta'(\sigma - \tilde{\sigma})
- {c \over 12\pi} \delta''' (\sigma - \tilde{\sigma})
\label{eq:2.25c}
\end{eqnarray}
\end{mathletters}%
where we have allowed for a possible central extension of strength $c$,
and it remains to calculate this quantity.

The gained advantage in two dimensional space-time is that all operators
are quadratic, see
(\ref{eq:2.19})-(\ref{eq:2.22});
the singularity structure may be assessed
and $c$ computed; obviously it is composed of independent contributions.
\beq
c = c^{\rm\,gravity} + c^{\rm\,matter} ~~,~~~~~~
c^{\rm\,gravity} = c_0 + c_1
\label{eq:2.26}
\eeq
Surprisingly, however,
there is more than one way of handling infinities and more than one
answer for $c$ can be gotten.  This reflects the fact, already known to
Jordan in the 1930s \cite{ref:15}, that an anomalous Schwinger term depends
on how the vacuum is defined.

In the present context, there is no argument about
$c^{\rm\,matter}$, the answer is
\beq
c^{\rm\,matter} = 1
\label{eq:2.27}
\eeq
The same holds for the positively signed gravity variable
(assume $\Lambda > 0$, so that $r^1$ enters positively).
\beq
c_1 = 1
\label{eq:2.28}
\eeq

But the negatively signed gravitational variable can be treated   in
more than one way, giving different answers for $c_0$.  The different
approaches may be named
``Schr\"odinger representation quantum field theory''
and ``BRST string/conformal field theory,''
and the variety arises owing to the various ways
one can quantize a theory with a negative kinetic term, like the $r^0$
gravitational variable.  (This variety is analogous to what is seen in
Gupta-Bleuler quantization of electrodynamics: the time component
potential $A_0$ enters with negative kinetic term.)

In the Schr\"odinger representation quantum field theory approach one
maintains positive norm states in a Hilbert space, and finds $c_0 = -1$,
$c^{\rm gravity} = c_0 + c_1 = 0$, $c = c^{\rm gravity} + c^{\rm matter}
= 1$.  Thus pure gravity has no obstructions, only matter provides the
obstruction.  Consequently the constraints of pure gravity can be solved,
indeed explicit formulas have been gotten by many people \cite{ref:16}.
In our formalism, according to
(\ref{eq:2.19})
and
(\ref{eq:2.20})
the constraints read
\begin{eqnarray}
{\cal E}^{\rm gravity} \big| \psi \big>_{\rm gravity} & \sim & {1 \over 2}
     \left({1 \over \Lambda} {\delta^2 \over \delta r^a \delta r_a} -
          \Lambda {r^a}' {r_a}' \right) \Psi_{\rm gravity} (r^a) = 0
\label{eq:2.29} \\
{\cal P}^{\rm gravity} \big| \psi \big>_{\rm gravity} & \sim & i {r^a}'
     {\delta \over \delta r^a} \Psi_{\rm gravity} (r^a) = 0
\label{eq:2.30}
\end{eqnarray}
with two solutions
\begin{mathletters}
\label{eq:2.31}
\begin{equation}
\Psi_{\rm gravity} (r^a) = {\rm exp} \pm i {\Lambda \over 2}
     \int d \sigma \epsilon_{ab} r^a {r^b}'
\label{eq:2.31a}
\end{equation}
This may also be presented by an action of a definite operator on the Fock
vacuum state $\big| 0 \big>$,
\begin{equation}
\Psi_{\rm gravity} (r^a) \propto \left[ \, {\rm exp} \pm \int dk \,
  {a_0}^{\!\dagger} (k) \, \epsilon (k) \,
  {a_1}^{\!\dagger} (-k) \right] \big| \, 0 \, \big>.
\label{eq:2.31b}
\end{equation}
with $  {a_a}^{\!\dagger} (k)$ creating field oscillations of definite
momentum.
\begin{equation}
{a_a}^{\!\dagger} (k)  = {-i \over \sqrt{4 \pi \Lambda |k|}}
\int d \sigma \, e^{i k \sigma} \, \pi_a (\sigma)
+ \sqrt{ {\Lambda |k| \over 4\pi} }
\int d \sigma \, e^{i k \sigma} \, r^a (\sigma)
\label{eq:2.31c}
\end{equation}
\end{mathletters}%
As expected, the state functional is invariant against
spatial coordinate redefinition, $\sigma \to \tilde{\sigma}(\sigma)$;
this is best seen by recognizing that integrand in the exponent of
(\ref{eq:2.31a})
is a 1-form:
$d\sigma \, \epsilon_{ab} \, r^a \, {r^b}' = \epsilon_{ab} \, r^a \, dr^b$.

Although this state is here presented
for a gravity model
in the Schr\"odinger
representation field theory context, it is also of interest to
practitioners of conformal field theory and string theory.
The algebra (\ref{eq:2.25}),
especially when written in decoupled form,
\begin{equation}
\Theta_{\pm} = {1 \over 2} ({\cal E} \mp P)
\label{eq:2.32}
\end{equation}
\vspace*{-0.3truein}
\begin{mathletters}
\label{eq:2.33}
\begin{eqnarray}
\left[ \Theta_{\pm} (\sigma), \Theta_{\pm} (\tilde{\sigma}) \right] & = &
  \pm i \left( \vphantom{1\over2}
\Theta_{\pm} (\sigma) + \Theta_{\pm} (\tilde{\sigma}) \right)
  \delta' (\sigma - \tilde{\sigma}) \mp {ic \over 24\pi} \delta'''
      (\sigma - \tilde{\sigma})
\label{eq:2.33a} \\
\left[ \Theta_{\pm} (\sigma), \Theta_{\mp} (\tilde{\sigma}) \right] & = & 0
\label{eq:2.33b}
\end{eqnarray}
\end{mathletters}%
is recognized as the position-space version of the Virasoro algebra and
the Schwinger term is just the Virasoro anomaly.  Usually one does not
find a field theoretic non-ghost realization {\it without\/} the Virasoro
center; yet the CGHS model, without matter provides an explicit example.
Usually one does not expect that {\it all\/} the Virasoro generators
annihilate a state, but in fact our states
(\ref{eq:2.31})
enjoy that property.

Once matter is added, a center appears, $c=1$, and the theory becomes
anomalous.  In the same Schr\"odinger representation approach used
above, one strategy is the following
modification of a method due to Kucha\v{r}
\cite{ref:14,ref:17}.  The Lagrange density (\ref{eq:2.16})
is presented in terms of decoupled constraints.
\begin{mathletters}
\begin{equation}
{\cal L} = \pi_a \dot{r}^a + \Pi \dot{\varphi} - \lambda^+ \Theta_+ - \lambda^-
\Theta_-
\label{eq:2.34a}
\end{equation}
\begin{equation}
\lambda^{\pm} = \alpha \pm \beta
\label{eq:2.34b}
\end{equation}
\end{mathletters}%
Then the gravity variables $\{ \pi_a, r^a \}$
are transformed by a linear canonical transformation to a new set
$\{ P_\pm, X^{\pm} \}$, in terms of which (2.34a) reads
\begin{mathletters}
\begin{equation}
{\cal L} = P_+ \dot{X}^+ + P_- \dot{X}^- + \Pi \dot{\varphi} - \lambda^+
     \left( P_+ {X^+}' + \theta_+^{\rm matter} \right) - \lambda^-
          \left(- P_- {X^-}' + \theta_-^{\rm matter} \right)
\label{eq:2.35a}
\end{equation}
\begin{equation}
\theta_{\pm}^{\rm matter} = {1 \over 4} (\Pi \pm \varphi')^2
\label{eq:2.35b}
\end{equation}
\end{mathletters}%
The gravity portions of the constraints $\Theta_{\pm}$
have been transformed to
$\pm P_{\pm} {X^{\pm}}'$ ---
expressions that look like momentum densities for fields
$X^{\pm}$, and thus satisfy the
$\Theta_\pm$ algebra
(\ref{eq:2.33})
without center, as do also momentum densities, see
(\ref{eq:2.12a}).

The entire obstruction in the full gravity plus matter constraints comes
from the commutator of the matter contributions $\theta_\pm^{\rm
matter}$.  In order to remove the obstruction, we modify the theory by
adding $\Delta \Theta_\pm$ to the constraint $\Theta_\pm$, such that no
center arises in the modified constraints.  An expression for
$\Delta \Theta_\pm$ that does the job is
\begin{equation}
\Delta \Theta_{\pm} = {1 \over 48 \pi} {({\rm ln} {X^{\pm}}')}''
\label{eq:2.36}
\end{equation}
Hence $\tilde{\Theta}_{\pm} \equiv \Theta_{\pm} + \Delta \Theta_{\pm}$
possess no obstruction in their algebra, and can annihilate states.
Explicitly, the
modified constraint equations read in the Schr\"odinger representation (after
dividing by ${X^{\pm}}'$)
\begin{equation}
\left( {1 \over i} {\delta \over \delta X^{\pm}} \pm {1 \over 48 \pi{X^{\pm}}'}
        \left({\rm ln} {X^{\pm}}'\right)'' \pm {1 \over { X^{\pm}}'}
         \theta_{\pm}^{\rm matter} \right) \psi  (X^{\pm}, \varphi) = 0
\label{eq:2.37}
\end{equation}
It is recognized that the anomaly has been removed by introducing
functional $U(1)$ connections in $X^{\pm}$ space, whose curvature
cancels the anomaly.  In the modified constraint there still is no
mixing between gravitational variables $\{ P_{\pm}, X^{\pm} \}$ and
matter variables $\{ \Pi, \varphi \}$.  But the modified gravitational
contribution is no longer quadratic --- indeed it is non-polynomial ---
and we have no idea how to solve
(\ref{eq:2.37}).
We suspect, however, that just as its matter-free version,
Eq.~(\ref{eq:2.37})
possesses only a few solutions ---
far fewer than the rich spectrum that emerges upon BRST quantization,
which we now examine.

In the BRST quantization method, extensively employed by string and
conformal field theory investigators, one adds ghosts, which carry their
own anomaly of $c_{\rm ghost} = -26$.  Also one improves $\Theta_\pm$ by
the addition of $\Delta\Theta_\pm$ so that $c$ is increased; for
example, with
\begin{eqnarray}
\Delta \Theta_\pm &=& {Q \over \sqrt{4\pi}} \left( \Pi \pm \varphi'
\right)'
\label{eq:2.38} \\
c &\to& c + 3 Q^2
\label{eq:2.39}
\end{eqnarray}
[The modification
(\ref{eq:2.38})
corresponds to ``improving''
the energy momentum tensor by
$(\partial_\mu \partial_\nu - g_{\mu\nu} \Box ) \varphi$].
The ``background charge'' $Q$ is chosen so that the total anomaly
vanishes.
\beq
c + 3Q^2 + c_{\rm ghost} = 0
\label{eq:2.40}
\eeq
Moreover, the constraints are relaxed by imposing that physical states
are annihilated by the ``BRST'' charges, rather than by the bosonic
constraints.  This is roughly equivalent to enforcing ``half'' the
bosonic constraints, the positive frequency portions.  In this way one
arrives at a rich and well known spectrum.

Within BRST quantization, the negative signed gravitational field $r^0$
is quantized so that negative norm states arise --- just as in
Gupta-Bleuler electrodynamics.   [Negative norm states cannot arise in a
Schr\"odinger representation, where the inner product is explicitly
given by a (functional) integral, leading to positive norm.]~
One then finds $c_0 = 1$; the center is insensitive to the signature
with which fields enter the action.   As a consequence, $c^{\rm gravity}
= c_0 + c_1 = 2$ so that even pure gravity constraints possess an
obstruction.

Evidently, pure gravity with $c^{\rm gravity} = 2$ requires $Q = 2
\sqrt{2}$.  The rich BRST spectrum is much more plentiful than the two
states
(\ref{eq:2.31})
found in the Schr\"odinger representation
and does not appear to reflect the fact that the classical pure gravity
theory is without excitations.

Gravity with matter carries $c=3$, and becomes quantizable at $Q =
\sqrt{23/3}$.  Once again a rich spectrum emerges, but it shows no
apparent relation to a flat-space particle spectrum.


We conclude that without question,
the CGHS model, and other similar two-dimensional
gravity models, are afflicted by anomalies in their constraint algebras,
which become second-class and frustrate straightforward quantization.
While anomalies can be calculated and are finite, their specific value
depends on the way singularities of quantum field theory are resolved,
and this leads to a variety of procedures for overcoming the problem and
to a variety of quantum field theories, with quite different properties.

Two methods were discussed:
(i) a Schr\"odinger representation with
Kucha\v{r}-type improvement
as needed, {\it i.e.\/} when matter is present,
and (ii) BRST quantization.
(Actually several other approaches are also available \cite{ref:14}.)
Only in the first method for pure matterless gravity,
with positive norm states and vanishing anomaly,
does the quantum
theory bear any resemblance to the classical theory,
in the sense that the classical gravity theory has no propagating
degrees of freedom, while the quantum Hilbert space has only
the two states in
(\ref{eq:2.31}),
neither of which contains any further degrees of
freedom.  In other cases,
{\it e.g.}~with matter, the classical picture of physics seems
irrelevant to the behavior of the quantum theory.

Presumably, if anomalies were absent, the different quantization
procedures (Schr\"odinger representation, BRST, $\ldots$) would
produce the same physics.  However, the anomalies {\it are\/} present and
interfere with equivalence.

Finally, we remark that our investigation has exposed an interesting
structure within Virasoro theory: there exists a field theoretic
realization of the algebra without the anomaly, in terms of spinless
fields and with no ghost fields.   Moreover, there are states that are
annihilated by {\it all\/} the Virasoro generators.

What does any of this teach us about the physical four-dimensional
model?  We believe that an extension in the constraint algebra will
arise for all physical, propagating degrees of freedom: for matter
fields, as is seen already in two dimensions, and also for gravity
fields, which in four dimensions (unlike in two) carry physical energy.
How to overcome this obstruction to quantization is unclear to us, but
we expect that the resulting quantum theory will be far different from
its classical counterpart.  Especially problematic is the fact that
flat-space calculations of anomalous Schwinger terms in four dimensions
yield infinite results, essentially for dimensional reasons.
Moreover, it should be clear that any
announced ``solutions'' to the constraints that result from
{\it formal\/} analysis must be viewed as preliminary:
properties of the Hilbert space and of the inner
product must be fixed first in order to give an unambiguous
determination of any obstructions.

We believe that our two-dimensional investigation, although in a much simpler
and unphysical setting, nevertheless contains important clues for
realistic theories.  Certainly that was the lesson of gauge theories:
anomalies and vacuum angle have corresponding roles in the Schwinger
model and in QCD!


\end{document}